\documentclass[11pt]{article}
\usepackage[margin=1.2in]{geometry}

\usepackage[utf8]{inputenc}
\usepackage[T1]{fontenc}
\usepackage{mathptmx}   
\usepackage[english]{babel}

\usepackage{authblk}      
\usepackage{titlesec}     
\usepackage{enumitem}     
\usepackage{csquotes}     
\usepackage{graphicx}     
\usepackage{xcolor}       

\usepackage{tikz}
\def\checkmark{\tikz\fill[scale=0.4](0,.35) -- (.25,0) -- (1,.7) -- (.25,.15) -- cycle;}

\usepackage[style=ieee, backend=biber]{biblatex}
\usepackage{hyperref}
\hypersetup{
    colorlinks=true,
    linkcolor=blue,
    filecolor=magenta,      
    urlcolor=cyan,
    citecolor=blue,
    pdftitle={The Conversational Exam},
    pdfauthor={Lorena A. Barba and Laura Stegner},
}

\title{\textbf{The Conversational Exam: A Scalable Assessment Design for the AI Era}}
\author[1]{Lorena A. Barba}
\author[2]{Laura Stegner}
\affil[1,2]{Mechanical and Aerospace Engineering Department, The George Washington University}
\date{}

\begin{document}

\maketitle

\begin{abstract}
Traditional assessment methods collapse when students use generative AI to complete work without genuine engagement, creating an illusion of competence where they believe they're learning but aren't. This paper presents the conversational exam—a scalable oral examination format that restores assessment validity by having students code live while explaining their reasoning. Drawing on human-computer interaction principles, we examined 58 students in small groups across just two days, demonstrating that oral exams can scale to typical class sizes. The format combines authentic practice (students work with documentation and supervised AI access) with inherent validity (real-time performance cannot be faked). We provide detailed implementation guidance to help instructors adapt this approach, offering a practical path forward when many educators feel paralyzed between banning AI entirely or accepting that valid assessment is impossible.
\end{abstract}

\section*{Introduction}
Two themes have dominated the narrative about generative artificial intelligence in education: the frothy hype from techno-optimists and corporations about ``revolutionizing'' education through personalization and on-demand intelligence, and the doomsday laments about students cheating and offloading their coursework to AI. Neither of these is helpful to managing the inevitable disruption. By most accounts, students are using AI extensively to get answers and to complete work for them. The question is how does higher education adapt to this reality, and ensure that learning still occurs in our classrooms.

One of us started 2024 post-sabbatical with full optimism, and embraced AI for teaching support in an undergraduate course in engineering computations. The course covers practical linear algebra with Python, numerical study of differential equations in simple settings like free fall and spring-mass systems, and a brief introduction to Fourier analysis of sound waves. It is the second in a two-course series, with the first being an introduction to programming with Python, exploratory data analysis, visualization and linear regression. Original lessons written for this course are open source on GitHub \cite{engineersCode_EngComp}, and the teaching method used in the classroom was the live-coding style popularized by the Software Carpentries workshops \cite{Wilson2014,Nederbragt2020TenTipsLiveCoding}, and favored by many educators in the scientific Python community. Homework assignments developed over several years used the \texttt{nbgrader} Jupyter extension \cite{Blank2019nbgrader} and a web-based system for autograding that gave students immediate feedback, with multiple attempts. This combination was not only successful but quite popular with students, who had many positive reactions, even in the worst of times during pandemic-era remote teaching. Then came generative AI, and a well-intentioned embrace of it by the instructor, with disastrous results. Students began using AI to copy and paste assignment questions directly, pasted AI-generated code solutions back in their Jupyter notebooks, and used the auto-grader via trial and error to get high marks without engaging. Attendance, satisfaction, and student learning all slumped.

In an account of that experience in this magazine’s Education Department \cite{Barba2025ExperienceGenAI}, the author highlighted one pernicious effect of students’ AI use, the \textit{illusion of competence} \cite{Koriat2005IllusionsCompetence}. It is a form of metacognitive miscalibration: in the presence of the answer, students misjudge their future performance, they have the \textit{feeling} that they are learning, but in fact they are not. The same can happen with other forms of passive engagement, like sitting in a lecture, or reading and highlighting notes or the textbook. But the effect is more severe with AI tools, because they so drastically increase a person’s capacity to complete certain tasks with minimal engagement.

Unfortunately, simply \textit{telling} students that they should not use AI for a task is ineffective. When unsupervised, they will use any assistance at hand, regardless of our instructions, our syllabus policy, or any ``traffic-light'' signals we add to homework assignments \cite{Corbin2025TalkIsCheap}. These discursive changes to assessment create an enforcement illusion, because it is impossible to reliably detect or verify students’ use of AI. Rules alone cannot protect assessment validity, and instead we should focus on \textit{structural} changes to assessment: redesigned tasks, interactive components, embedded process evidence, and so on.

This paper describes one assessment design that we developed for addressing the challenge of validity in the presence of AI: the \textit{conversational exam}. It came about from pondering that examination of each student in an oral setting with an instructor is a fail-proof way to see what they know. In large classes, however, oral exams become impractical (our class had close to 60 students). A spark of inspiration came from seeing some media commentary on tech companies using group interviews, where several candidates join a video call simultaneously. With some trepidation, a sentence on the syllabus sealed our commitment for this new exam format, and we embarked on its design. We hope that the detailed description here will help other instructors who wish to adapt the concept for their contexts.

\section*{Design Principles and Rationale}

\subsection*{Core Design Principles}
The conversational exam rests on three ideas as the foundation: (1) assessment should ideally mirror authentic practice, (2) validity should derive from the format itself rather than from policing student behavior, and (3) oral examination can be made practical through careful attention to logistics and cognitive load. Let’s examine these ideas.

\begin{description}
    \item[Authenticity over artificial constraints.] In professional technical work, people code with access to documentation, online resources, and increasingly, AI tools. They collaborate with colleagues, search for solutions to error messages, and reference examples from past projects. Creating exam conditions that strip away all these supports—locked-down browsers, banned tools, isolated workspaces—may produce valid measurements of what students can do in those constrained circumstances, but it tells us little about their readiness for the actual work of computing. The conversational exam takes a different approach. Students work in their familiar environment, a cloud-based Jupyter notebook, with documentation and course materials at hand, and connected to the open internet. They may even use AI, though under supervision and with clear boundaries about what constitutes appropriate use. The assessment focuses on what they can do and explain in real time, with the time constraint being perhaps the one ``exam condition'' that deviates from work scenarios.
    
    \item[Validity through performance, not prohibition.] Recent scholarship on assessment in the age of AI has called for a shift in perspective: rather than obsessing over student behavior (are they cheating?), we should focus on the properties of the assessment itself (does it validly measure learning?) \cite{Dawson2024ValidityMattersMore}. An assessment that depends on students \textit{not using} AI is fragile; its validity collapses the moment a student finds a way around our barriers. The conversational exam builds validity into its structure. Live coding with concurrent explanation cannot be faked through memorization or delegated to AI. When a student types code while articulating their reasoning, pauses to debug an error, and responds to follow-up questions about why their approach works, they reveal their actual understanding. A student who has relied on AI to complete coursework will stumble when asked to explain their process and write code on the fly. The performance itself becomes the evidence, and with appropriate guidance in advance, students recognize the need to prepare and engage with the practice opportunities we provide them.
    
    \item[Scalability through thoughtful design.] The conventional wisdom says that oral examinations don't scale, and for good reason: examining 58 students individually could take days of instructor time. Instead, the conversational exam assesses students in small groups of five or six, taking turns to answer questions while the others observe. This rotation keeps everyone engaged, as students watch their peers code and learn from the questions asked. It diffuses some of the anxiety that comes with one-on-one examination, creating a more conversational and less inquisitorial atmosphere. The group format also allows for some operational efficiency, as we describe below. It requires some task allocation to work well: the lead instructor guides the questioning, the co-instructor marks students’ performance on a pre-prepared sheet while they code and explain, and the teaching assistant monitors technical issues and helps keep time. What seemed impossible—validly assessing a full class through oral examination—becomes entirely feasible. We examined 58 students across 10 half-hour sessions.
\end{description}

\subsection*{HCI-Informed Process Design}
The design also draws on principles from human-computer interaction (HCI) research, particularly around cognitive-load management and consistency in evaluation. One of us has extensive experience conducting user studies, where careful preparation is essential to managing the heavy multitasking demands of live observation and note-taking, which are inherently mentally demanding. Past research has found that interview quality is decreased if the interviewer’s working memory is overloaded \cite{Giorgianni2025InterviewerCognitiveLoad}. We applied those same principles here.

Each question comes with a detailed marking sheet: a grid for checking off rubric items, the expected solution, a hint if the student struggles, an elevated concept check if they succeed quickly, and a simplified fallback question. This approach follows established best practices for structured observation, which emphasize that evidence must be gathered and recorded systematically, rather than through happenstance, to serve as a dependable index of performance \cite{Croll1986SystematicClassroomObservation}. In HCI methodology, such a documented process acts as an externalized artifact that establishes a \textit{shared conception} of effectiveness among the team; this standardizes how different observers evaluate behavior to increase consistency in assessment \cite{Hartson2018UXBook}. Furthermore, the use of clear, pre-determined question and answer flows mirrors the rigorous ``behavior instructions'' required in Wizard of Oz prototyping \cite{Kelley1984IterativeDesignWizardOfOz} where a human operator discretely controls the behavior of a computer or robotic system to simulate an autonomous system. In Wizard of Oz methodology, maintaining a deterministic system logic through unambiguous operator instructions is paramount for data integrity; without this structure, investigations can suffer from random effects or unintended improvisation that undermines the evaluation. Finally, from HCI field work experience, we worked hard to fit everything \textit{on one page}, to eliminate the need for paper shuffling during the observation. This preparation and organization shifts cognitive burden from the moment of examination to the work done beforehand. During the exam itself, the instructor can focus entirely on the student's performance rather than scrambling to invent follow-up questions or remember grading criteria. The structure provides flexibility—we can adapt based on how each student performs—without requiring improvisation under pressure.

\subsection*{Question Development and Administration}
We built a question bank for the exam in two tiers of difficulty, with 30 questions in each tier. The first tier tests foundational skills: the building blocks, the syntax and vocabulary of computational work. The second tier probes more conceptual understanding or requires problem-solving skills in context. Each question targets a specific learning objective—say, computing eigenvalues or normalizing a vector—and can be completed in a minute or two by a prepared student.

In each round, students draw a question from a paper bag (one for each tier), which creates both physical randomization and a bit of theatrical suspense. The student is asked to state the question number (so we can find the corresponding marking sheet in our pre-organized stack) and then read the question out loud, at which point we turn over a toy sand clock for timekeeping.

\begin{center}
\fbox{\begin{minipage}{0.6\textwidth}
\textbf{Three-Tier Question Scaffolding}
\begin{itemize}
    \item \textbf{Level 1 (2 minutes):} Foundational recall and basic syntax—minimal competence
    \item \textbf{Level 2 (3 minutes):} Conceptual application—shows deeper understanding
    \item \textbf{Optional Level 3:} Problem-solving integration for high-performing students
\end{itemize}
\end{minipage}}
\end{center}

The real innovation lies in how we prepared for each question. Rather than writing just the question text and expected answer, we developed a decision tree. If a student succeeds quickly, we have an elevated concept check ready: e.g., ``What would happen if you used \texttt{axis=1} instead?'' or ``How would this change if the matrix weren't symmetric?'' These probing questions reveal whether the student truly understands the concept or simply memorized a procedure (and shift our marking of their conceptual understanding from ``proficient'' to ``excellent''). If a student struggles, we can offer a hint without improvising: e.g., ``Remember that \texttt{numpy.linalg.norm()} takes an axis parameter.'' If they continue to flounder, we have a simplified version on hand: e.g., ``instead of normalizing multiple vectors, normalize just one'' (and we shift the marking down to ``developing''). This scaffolding ensures that every student, regardless of which question they draw, has multiple opportunities to demonstrate their competence. Nobody faces a dead end in awkward silence.

Creating the question bank took considerable effort upfront—and indeed the assistance of generative AI tools was valuable in this stage. We mapped the course learning objectives and topics to specific assessable tasks, paying attention to what could reasonably be demonstrated in a few minutes. Some topics lend themselves naturally to this format—vector operations, matrix manipulations, basic function calling—while others require more careful thought. For instance, we cannot ask students to implement the full singular value decomposition algorithm in three minutes, but we can ask them to use \texttt{numpy.linalg.svd()} and explain what the singular values represent. The constraint of time, rather than being a limitation, forced us to focus on what truly matters: can the student apply the concept and articulate why it works? Our attention to detail in the preparation stage made the actual examination far less cognitively demanding for the instructional team, even as we conducted ten consecutive sessions.

\section*{Implementation: Making It Work}

\subsection*{Logistical Setup}
The conversational exam requires coordination among people, technology, and physical materials. We had a team of three: the lead instructor who conducts the questioning, a co-instructor who handles the marking sheets and monitors screens, and a teaching assistant who manages time and handles technical issues like choosing the student screen to view on the projector. This division of labor proved essential. Trying to question a student, mark their performance, watch for AI misuse, and track time simultaneously would overwhelm any single person.

For scheduling, we used \texttt{zcal}, \footnote{https://zcal.co} a web-based appointment-booking tool that integrates with calendar systems. Students selected their preferred 30-minute slot from available times, and the system automatically sent calendar invitations to them and to the examination team. This self-service approach eliminated the back-and-forth of coordinating schedules with 58 people. We offered slots across two days, with a break for lunch.

The room setup was augmented with Zoom, which turned out to have a feature we hadn't known about: multiple participants can share their screens simultaneously \cite{ZoomSupportMultiShare}. This capability was very useful. All five or six students in each group joined the same Zoom meeting, while seated together at a large table with their laptops open. Everyone joined with microphones muted by default and shared their full screen. When it was a student's turn to answer, the TA switched to showing that screen on his laptop connected to a projector, so we could all watch them code in real time. The co-instructor monitored a large external display that showed a few shared screens in popped-out windows and others on tabs, which allowed for quick detection if a student opened an AI chat window or navigated away from their Jupyter notebook. Students worked in JupyterHub, the cloud-based notebook environment we use throughout the course, with access to documentation and course materials at will.

\subsection*{Examination Flow}
Each 30-minute session followed a predictable rhythm, which helped manage both our cognitive load and student anxiety. The first few minutes were a warm-up: all students simultaneously would open a Jupyter notebook, import NumPy and Matplotlib, and run a simple command. This settled nerves, and confirmed that everyone's technical setup worked. After the warm-up, we moved into the first rotation of questions. In turn, each student drew a card from the Level 1 bag, announced their question number, and read the question aloud. The TA turned the sandclock, and the student would then begin typing in their Jupyter notebook while explaining their thinking out loud. A question like ``Create a NumPy array with ten equally spaced numbers from zero to pi, and compute its magnitude'' takes less than a minute for a prepared student. The lead instructor would watch and listen, offer the hint if a student struggled, or move to the elevated concept check if time allowed. The co-instructor would be checking boxes on the marking sheet and noting whether the student explained clearly, made errors, needed hints, or sailed through confidently. The TA tracked time and managed the pre-organized stack of solution and marking sheets.

\begin{center}
\fbox{\begin{minipage}{0.8\textwidth}
\textbf{Three-Category Grading Rubric (1-4 scale, no zeros)}
\begin{itemize}
    \item \textbf{Technical Skills (10 points):} Python syntax, NumPy functions, code execution, error handling
    \item \textbf{Conceptual Understanding (10 points):} Mathematical explanation, geometric interpretation, concept connections, result verification
    \item \textbf{Problem-Solving \& Communication (5 points):} Logical approach, debugging ability, response to hints, explanation clarity
\end{itemize}
\end{minipage}}
\end{center}

The marking sheets were the linchpin of the whole operation. Each sheet had the question text at the top, the expected solution with code, space for annotations, the hint/elevated/simplified question variants, and a rubric grid with checkboxes (technical skills 1-4, conceptual understanding 1-4, problem-solving 1-4) at the bottom. During the exam, the co-instructor would mark the grid in real time, adding quick notes about what the student did or said. To generate a final score out of 100, raw ratings from the 1–4 forced-choice scale were transformed using criterion-specific multipliers. Technical competency and conceptual understanding were each assigned a multiplier of 10 (max 40 points each), while problem-solving was assigned a multiplier of 5 (max 20 points). This weighting strategy ensured that the assessment prioritized core technical and mathematical mastery. Individual totals from the two questions were then averaged to produce the final numerical grade.

The second rotation used the Level 2 bag, with slightly harder questions that required more conceptual understanding. By this point, about 15 minutes into the session, students had watched several peers answer questions and had seen the format in action. The anxiety level had hopefully dropped. Some students who were tentative at the start would be leaning in, watching intently, perhaps making connections between what they saw and what they might be asked. The format has a strange alchemy: it was an exam, high-stakes and graded, but it felt more like a seminar where students were demonstrating their work to their peers.

Throughout the session, we tried to maintain an encouraging tone. When a student typed something correct, we would nod or say ``Good, keep going.'' When they made a small error, we might say ``Check your indexing there'' rather than letting them spiral into confusion. The goal was to see what they could do with minimal support, but we weren't interested in watching anyone fail spectacularly. This conversational approach—genuinely treating it as a conversation rather than an interrogation—made a difference in how students performed. They were more willing to think aloud, to say ``I'm not sure about this part,'' and to recover from small mistakes.

The routine became almost automatic by the third or fourth session, which was precisely the point. The heavy cognitive lifting had been done during preparation, leaving us free to focus entirely on student performance during the exam itself. The meticulously prepared exam sheets enabled us to maintain a confident, rapid pace throughout the day, counteracting the observer fatigue that can cause individual sessions and answers to blur together. Immediately after a day’s sessions, while our memories were fresh, the three of us would do a quick debrief, flagging any students who seemed particularly strong or weak. We would then convert those checkbox marks into numerical scores and write brief comments if needed. After the two days of examination, we were able to post the grades within just hours.

\section*{Outcomes and Reflection}
As each group of students poured into the room, we saw them arrange their hand-written notes on the desk, or have digital notes on another window open on their laptop. They had prepared. On several occasions they asked us if they could enter an error message into ChatGPT, or Google something (which inevitably at this point brings up answers in AI Mode). We allowed it, mindful of the AI use policy for the exam that we had shared with them in advance. As the sessions progressed one after the other, it was quite clear who had meaningfully engaged with the course, and who had not. Some students were nervous, of course, but the structured format allowed them to perform well regardless (if prepared). In the end, the class average on the exams (two instances) was around 80 percent. The full experience reassured us: it works!

\begin{center}
\fbox{\begin{minipage}{0.7\textwidth}
\textbf{AI Use Policy for Conversational Exam}

\textit{Permitted AI Use:}
\begin{itemize}[label=\checkmark]
    \item Explaining error messages after code fails
    \item Clarifying Python documentation
    \item Understanding why their code didn't work (post-attempt)
    \item Explaining concepts from their own notes
\end{itemize}

\textit{Prohibited AI Use:}
\begin{itemize}[label=$\times$]
    \item Generating code for them
    \item Copy-pasting AI-generated code
    \item Asking ``how do I solve this problem'' before attempting
    \item Using AI to write functions, loops, or logic
\end{itemize}
\end{minipage}}
\end{center}

Not all is rosy, however. For many students it was an alien experience, despite receiving written guidance about the exam format to know what to expect, and having a class activity where we directed them to use Google Gemini in Guided Learning mode, in pairs, to practice answering while live-coding. A few who answered a question poorly were very frustrated, and perceived the randomized questions as ``unfair''—perhaps this always happens with exams, but they identified the novel format as the culprit. We believe that over time, if oral exams gain more traction as a countermeasure to AI misuse, a culture shift will take place.

\section*{Summing-up The Conversational Exam}
The experience of, first, \textit{deciding} to implement an authentic assessment that evaluates students person-to-person, then \textit{designing} a format that could scale to (moderately) large groups, and finally orchestrating it all, was intense. The decision was made with trepidation, the design was laborious, and the orchestration daunting for all involved. We encountered some unanticipated pitfalls: for example, how to handle students with extra-time accommodations? Aware that we needed to be discreet and not disclose a student’s accommodation to the others in a group, we decided to use the same timers for all, but the instructor doing the grading would inconspicuously observe the student with accommodations for an extra minute or two. (Nearly all students tended to continue working on their question after their time was up and the projector had been switched to sharing another’s screen.) This tactic seemed to work during the first exam run, but one student complained. After discussing it with them, it became apparent that the student preferred to plainly \textit{see} their extra time being granted. Thus, for the second run we had to individually get each student with time accommodations to agree ahead of time to us using a visibly different timer for them. We had several calls with staff at the disability services office, and it was time consuming. In the future, we will explore formally applying for a ``fundamental alteration,'' which gives an exception for cases when a requested accommodation fundamentally alters an essential requirement of a course. This is pending.

Despite the stumbling blocks, however, we are satisfied with the results. The conversational exam offered a valid form of assessment even in the presence of generative AI, made students feel accountable for their learning, and felt authentic to the work settings where students might apply their knowledge in the future. Below we summarize the merits of the conversational exam.

\begin{description}
    \item[Promoting Deep Mastery via Adaptive Probing:] By probing the ``edges'' of a student's mental model, this format shifts student study strategies away from procedural memorization toward conceptual mastery. The environment allows for ``adaptive probing,'' where structured follow-up questions help instructors distinguish superficial recall from a student's true ability to apply knowledge to real-world scenarios.
    
    \item[Ensuring Resilient Integrity in the Age of AI:] As generative AI makes traditional written and procedural assignments increasingly difficult to verify, the conversational format provides a primary defense for academic integrity. By requiring real-time, unscripted reasoning and the spontaneous defense of one's logic, the format makes it inherently difficult to outsource thinking to AI and ensures that students demonstrate genuine intellectual ownership.
    
    \item[Scalability via Organizational Creativity and HCI Rigor:] By applying HCI-informed organizational design—such as ``Wizard of Oz'' deterministic logic and meticulous marking sheets—instructors can overcome the logistical barriers and high labor costs historically associated with oral exams. This ``scalability through structure'' enables high-fidelity assessment to be deployed across larger cohorts efficiently, shifting the heavy cognitive lifting to the preparation phase and ensuring that human examiners maintain consistency without the risk of unintended improvisation or observer fatigue.
\end{description}

\section*{Closing Thoughts}
The conversational exam emerged from necessity—a response to watching traditional assessment crumble in the face of students' unfettered AI use. It began as an act of desperation, then became something more valuable: a demonstration that we can design assessment for the world as it actually is, rather than the world we wish still existed. The format acknowledges that AI tools are ubiquitous and increasingly powerful, yet refuses to capitulate to the notion that authentic evaluation is impossible in their presence.

We made mistakes, encountered unanticipated problems, and spent countless hours refining details that seemed trivial until they became critical. But the format worked, and more importantly, it scaled. Fifty-eight students examined in two days, with valid results that differentiated genuine competence from surface performance. It was a proof of concept that valid assessment remains achievable when we attend carefully to structure and process.

The broader challenge remains daunting. Every semester, students graduate having relied on AI to complete their coursework without developing genuine competence. Every semester, institutions delay adapting their assessment practices, hoping perhaps that the AI disruption will somehow resolve itself or fade away. It won't. The technology is advancing too rapidly, becoming too integrated into every aspect of knowledge work. The question is whether education will lead this transformation or merely react to it.

We share this work not as a panacea, but as evidence that alternatives exist. The conversational exam suits computational courses particularly well, but the underlying principles—authenticity, inherent validity, scalable structure—apply more broadly. Other disciplines will need their own innovations, adapted to their contexts and constraints. What matters is that we begin experimenting, documenting outcomes honestly, and building collective knowledge about what works. The paralysis many instructors feel comes from seeing only two paths: ban AI entirely or accept that valid assessment is impossible. There are other paths and we need to find them together.

The conversational exam taught us something beyond its immediate practical value. It reminded us that meaningful assessment requires human judgment and presence in ways that cannot be automated or outsourced. The format succeeds because it centers on conversation—on the irreducibly human exchange between an instructor who knows the terrain and a student navigating it for the first time. That exchange, properly structured and thoughtfully conducted, creates a space where genuine understanding becomes visible. Perhaps this is what assessment should have been all along: not an exercise in surveillance and ranking, but an opportunity for students to demonstrate their competence and for instructors to witness it. We might emerge from this disruption with better assessment practices than we had before, if we have the courage to experiment and the humility to learn from our failures.

\section*{Author Bios}
\noindent\textbf{Lorena A. Barba} is a professor of mechanical and aerospace engineering at the George Washington University. She is a member of the IEEE Computer Society and past EiC of \textit{Computing in Science and Engineering}. Contact her at \href{mailto:labarba@gwu.edu}{labarba@gwu.edu}

\bigskip

\noindent\textbf{Laura Stegner} is an assistant professor of mechanical and aerospace engineering at the George Washington University in the area of robotics. Contact her at \href{mailto:laura.stegner@gwu.edu}{laura.stegner@gwu.edu}

\newpage
\appendix
\section{Implementation Checklist for Adopters}

\textbf{Pre-Exam Preparation (2-3 weeks before):}
\begin{itemize}
    \item Design question bank (30-60 questions depending on rounds)
    \item Create marking sheets for each question
    \item Develop rubric and calibrate with co-instructors
    \item Set up scheduling system (\texttt{zcal} or similar)
    \item Communicate format and expectations to students
    \item Conduct practice/sample question session with students
\end{itemize}

\textbf{One Week Before:}
\begin{itemize}
    \item Finalize question selection and print question cards
    \item Test technical setup (Zoom screen-sharing, monitor visibility)
    \item Prepare physical materials (paper bag, marking sheet packets)
    \item Send reminders to students with technical setup instructions
\end{itemize}

\textbf{Exam Day:}
\begin{itemize}
    \item Test technical setup 30 minutes before first session
    \item Brief examination team on roles and protocols
    \item Maintain consistent energy and encouraging tone
    \item Debrief between sessions for calibration
\end{itemize}

\textbf{Post-Exam:}
\begin{itemize}
    \item Complete scoring within 24 hours (while fresh)
    \item Calibrate any borderline cases with team
    \item Document lessons learned for future iterations
\end{itemize}

\newpage

\section{Example of a fully scaffolded question sheet used during the exam.}
\begin{figure}[htbp]
  \centering
  \includegraphics[width=0.9\columnwidth]{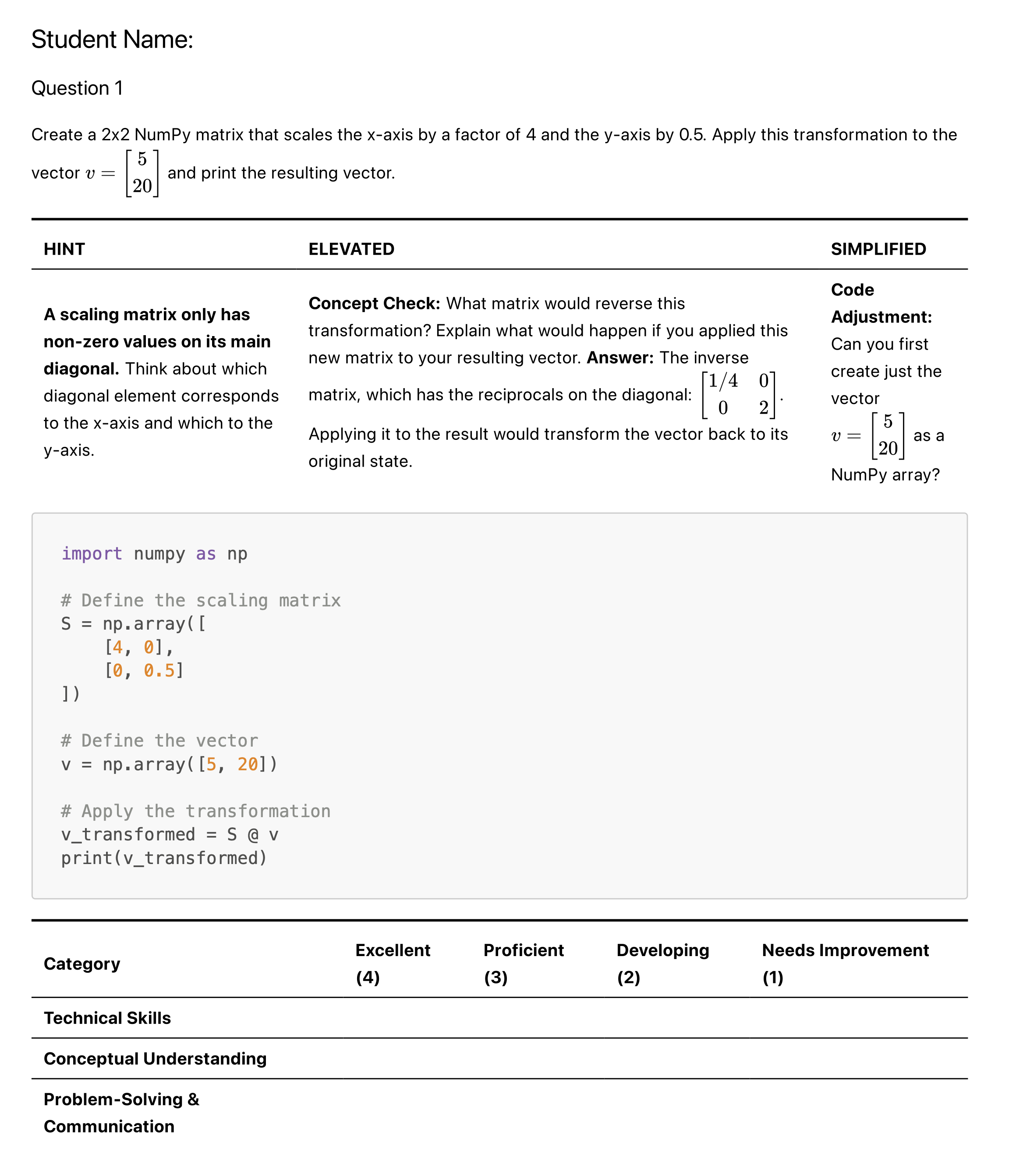}
  \label{fig:questionsheet}
\end{figure}

\clearpage
\printbibliography

@article{Wilson2014,
  title = {Software {Carpentry}: lessons learned},
  author = {Wilson, Greg and Aruliah, D. A. and Brown, C. Titus and Chue Hong, Neil P. and Davis, Matt and Guy, Richard T. and Haddock, Steven H. D. and Huff, Kathryn D. and Mitchell, Ian M. and Plumbley, Mark D. and Waugh, Ben and White, Ethan P. and Wilson, Paul},
  journal = {F1000Research},
  volume = {3},
  year = {2014},
  publisher = {Faculty of 1000 Ltd},
  doi = {10.12688/f1000research.3-62.v2}
}

@online{nbgrader,title = {nbgrader: A system for assigning and grading notebooks},author = {{Project Jupyter}},year = {2024},url = {https://nbgrader.readthedocs.io/},urldate = {2024-01-01}}

@misc{engineersCode_EngComp,
  author       = {Barba, Lorena A.},
  title        = {Engineering Computations ({EngComp}) course materials},
  howpublished = {\url{https://github.com/engineersCode/EngComp}},
  note         = {Open-source lessons for engineering computations},
  year         = {2017}
}

@article{Nederbragt2020TenTipsLiveCoding,
  author  = {Nederbragt, Alexander and Harris, Richard M. and Hill, April P. and Wilson, Greg},
  title   = {Ten quick tips for teaching with participatory live coding},
  journal = {PLOS Computational Biology},
  year    = {2020},
  volume  = {16},
  number  = {9},
  pages   = {e1008090},
  doi     = {10.1371/journal.pcbi.1008090}
}

@article{Blank2019nbgrader,
  author  = {Blank, Douglas S. and Bourgin, David and Brown, Allison and Bussonnier, Matthias and Frederic, Jeremy and Granger, Brian and Griffiths, Thomas L. and Hamrick, Jessica and Kelley, Kyle and Pacer, M. and Page, L.},
  title   = {nbgrader: A tool for creating and grading assignments in the {Jupyter Notebook}},
  journal = {The Journal of Open Source Education},
  year    = {2019},
  volume  = {2},
  number  = {11},
  doi     = {10.21105/jose.00032}
}

@article{Barba2025ExperienceGenAI,
  author  = {Barba, Lorena A.},
  title   = {Experience Embracing {GenAI} in an Engineering Computations Course: What Went Wrong and What’s Next},
  journal = {Computer},
  year    = {2025},
  volume  = {58},
  number  = {8},
  pages   = {136--141},
  doi     = {10.1109/MC.2025.3572654}
}

@article{Koriat2005IllusionsCompetence,
  author  = {Koriat, Asher and Bjork, Robert A.},
  title   = {Illusions of competence in monitoring one's knowledge during study},
  journal = {Journal of Experimental Psychology: Learning, Memory, and Cognition},
  year    = {2005},
  volume  = {31},
  number  = {2},
  pages   = {187--194},
  doi     = {10.1037/0278-7393.31.2.187}
}

@article{Corbin2025TalkIsCheap,
  author  = {Corbin, Tom and Dawson, Phillip and Liu, Di},
  title   = {Talk is cheap: why structural assessment changes are needed for a time of {GenAI}},
  journal = {Assessment \& Evaluation in Higher Education},
  year    = {2025},
  volume  = {50},
  number  = {7},
  pages   = {1087--1097},
  doi     = {10.1080/02602938.2025.2503964}
}

@article{Dawson2024ValidityMattersMore,
  author  = {Dawson, Phillip and Bearman, Margaret and Dollinger, Mollie and Boud, David},
  title   = {Validity matters more than cheating},
  journal = {Assessment \& Evaluation in Higher Education},
  year    = {2024},
  volume  = {49},
  number  = {7},
  pages   = {1005--1016},
  doi     = {10.1080/02602938.2024.2386662}
}

@article{Giorgianni2025InterviewerCognitiveLoad,
  author  = {Giorgianni, Davide and Vrij, Aldert and Leal, Sharon and Deeb, Haneen},
  title   = {The Effect of the Interviewer’s Cognitive Load on the Quality of the Forensic Interview},
  journal = {The European Journal of Psychology Applied to Legal Context},
  year    = {2025},
  volume  = {17},
  number  = {2},
  pages   = {101--110},
  doi     = {10.5093/ejpalc2025a9}
}

@book{Croll1986SystematicClassroomObservation,
  author    = {Croll, Paul},
  title     = {Systematic Classroom Observation},
  series    = {Social Research and Educational Studies},
  volume    = {3},
  publisher = {Taylor \& Francis},
  address   = {New York},
  year      = {1986},
  pages     = {202}
}

@book{Hartson2018UXBook,
  author    = {Hartson, Rex and Pyla, Pardha S.},
  title     = {The UX Book: Agile UX Design for a Quality User Experience},
  edition   = {2},
  publisher = {Morgan Kaufmann},
  year      = {2018}
}

@article{Kelley1984IterativeDesignWizardOfOz,
  author  = {Kelley, John F.},
  title   = {An iterative design methodology for user-friendly natural language office information applications},
  journal = {ACM Transactions on Information Systems},
  year    = {1984},
  volume  = {2},
  number  = {1},
  pages   = {26--41}
}

@misc{ZoomSupportMultiShare,
  author       = {{Zoom Video Communications, Inc.}},
  title        = {Enabling multiple participants to share their screens simultaneously},
  howpublished = {\url{https://support.zoom.com/hc/en/article?id=zm_kb&sysparm_article=KB0064237}},
  note         = {Accessed: 2026-01-15},
  year         = {2024}
}

\end{document}